\documentclass{amsart}
\usepackage{amssymb, latexsym}

\newcommand{\beqa}{\begin{eqnarray*}}
\newcommand{\eeqa}{\end{eqnarray*}}
\newcommand{\beqn}{\begin{eqnarray}}
\newcommand{\eeqn}{\end{eqnarray}}

\newcommand{\iy}{\infty}

\newcommand{\Ha}{\mathbb H}

\newcommand{\f}{\frac}

\newcommand{\al}{\alpha}

\newcommand{\om}{ \omega}
\newcommand{\Om}{\Omega}

\newcounter{cnt1}
\newcounter{cnt2}
\newcounter{cnt3}
\newcommand{\blr}{\begin{list}{$($\roman{cnt1}$)$}
 {\usecounter{cnt1} \setlength{\topsep}{0pt}
 \setlength{\itemsep}{0pt}}}
\newcommand{\bla}{\begin{list}{$($\alph{cnt2}$)$}
 {\usecounter{cnt2} \setlength{\topsep}{0pt}
 \setlength{\itemsep}{0pt}}}
\newcommand{\bln}{\begin{list}{$($\arabic{cnt3}$)$}
 {\usecounter{cnt3} \setlength{\topsep}{0pt}
 \setlength{\itemsep}{0pt}}}
\newcommand{\el}{\end{list}}
\newtheorem{thm}{Theorem}

\newtheorem{cor}[thm]{Corollary}
\newtheorem{ex}[thm]{Example}

\newtheorem{Def}[thm]{Definition}

\newtheorem{rem}[thm]{Remark}
\newcommand{\Rem}{\begin{rem} \rm}
\newcommand{\bdfn}{\begin{Def} \rm}
\newcommand{\edfn}{\end{Def}}

\newcommand{\ba}{\begin{array}}
\newcommand{\ea}{\end{array}}

\newtheorem*{acknowledgements}{Acknowledgements}
\date{}
\begin{document}
\title{\bf Sufficiency Class for Global (in time) Solutions to The 3D-Navier-Stokes Equations}
\author[Gill]{T. L. Gill}
\address[Tepper L. Gill]{ Department of Electrical Engineering, Howard University\\
Washington DC 20059 \\ USA, {\it E-mail~:} {\tt tgill@howard.edu}}
%\and
\author[Zachary]{W. W. Zachary}
\address[Woodford W. Zachary]{ Department of Electrical Engineering \\ Howard
University\\ Washington DC 20059 \\ USA, {\it E-mail~:} {\tt
wwzachary@earthlink.net}}
% [-1ex] \normalsize \sc Howard University\\
%\abstract{ write abstract here}
\date{}
%thispagestyle{empty}
\begin{abstract}  Let $ \Om $ be an open domain of class $\mathbb{C}^2 $ contained in ${\mathbb {R}}^3 $, let ${L^2(\Om)}^3$ be the Hilbert space of square integrable functions on ${ \Om} $ and let ${\mathbb H}[ \Om]:=\Ha$ be the completion of the set, $\left\{ {{\bf{u}} \in (\mathbb {C}_0^\infty  [ \Om ])^3 \left. {} \right|\,\nabla  \cdot {\bf{u}} = 0} \right\}$, with respect to the inner product of ${L^2(\Om)}^3$.  A well-known unsolved problem is the construction of a sufficient class of functions in ${\mathbb H}$ which will allow global, in time, strong solutions to the three-dimensional Navier-Stokes equations.  These equations describe the time evolution of the fluid velocity and pressure of an incompressible viscous homogeneous Newtonian fluid in terms of a given initial velocity and given external body forces.  In this paper, we use the analytic nature of the Stokes semigroup to construct an equivalent norm for $\mathbb{H}$, which provides strong bounds on the nonlinear term.  This allows us to prove that, under appropriate conditions, there exists a number $ {{{u}}_+} $, depending only on the domain, the viscosity, the body forces and the eigenvalues of the Stokes operator, such that, for all functions in a dense set $\mathbb{D}$ contained in the closed ball ${{\mathbb B} ( \Om )}=:{\mathbb B}$ of radius $ \tfrac{1}{2}{{u}_ +} $ in ${{\mathbb H}}$, the Navier-Stokes equations have unique, strong, solutions in ${\mathbb C}^{1} \left( {(0,\infty ),{\mathbb H}} \right)$.
\end{abstract}
\maketitle
\section*{Introduction} Let $ \Om $ be an open domain of class $\mathbb{C}^k $ contained in $\mathbb{R}^n $, $ n \ge 2 $, let $({L}^2[ \Om ])^n$ be the Hilbert space of square integrable functions on $ \Om $ with values in $\mathbb{C}^n $, let $\mathbf{D}[ \Om]$ be 
$\{ {\mathbf{u}} \in (\mathbb{C}_0^\infty  [ \Om ])^n \left. {} \right|\,\nabla  \cdot {\mathbf{u}} = 0\} $, let $\mathbb{H}$ be the completion of $\mathbf{D} [ \Om] $ with respect to the inner product of 
$({L}^2 [ \Om ])^n $, and let $\mathbb{V}[ \Om ]$ be the completion of $\mathbf{D} [ \Om]$ with respect to the inner product of $\mathbb{H}^1 [ \Om ]$, the functions in $\mathbb{H}$ with weak derivatives in $({L}^2 [ \Om ])^n $.  

The global in time classical Navier-Stokes initial-value problem (for $\Omega  \subset \mathbb{R}^n ,{\text{ and all }}T > 0$) is to find functions ${\mathbf{u}}:[0,T] \times \Omega  \to \mathbb{R}^n,$ and $p:[0,T] \times \Omega  \to \mathbb{R}$, such that
\beqn
\begin{gathered}
  \partial _t {\mathbf{u}} + ({\mathbf{u}} \cdot \nabla ){\mathbf{u}} - \nu \Delta {\mathbf{u}} + \nabla p = {\mathbf{f}}(t){\text{ in (}}0,T) \times \Omega , \hfill \\
  {\text{                              }}\nabla  \cdot {\mathbf{u}} = 0{\text{ in (}}0,T) \times \Omega , \hfill \\
  {\text{                              }}{\mathbf{u}}(t,{\mathbf{x}}) = {\mathbf{0}}{\text{ on (}}0,T) \times \partial \Omega , \hfill \\
  {\text{                              }}{\mathbf{u}}(0,{\mathbf{x}}) = {\mathbf{u}}_0 ({\mathbf{x}}){\text{ in }}\Omega . \hfill \\ 
\end{gathered} 
\eeqn
The equations describe the time evolution of the fluid velocity ${\mathbf{u}}({\mathbf{x}},t)$ and the pressure $p$ of an incompressible viscous homogeneous Newtonian fluid with constant viscosity coefficient $\nu $ in terms of a given initial velocity ${\mathbf{u}}_0 ({\mathbf{x}})$ and given external body forces
${\mathbf{f}}({\mathbf{x}},t)$.

The existence of global weak solutions of (1) was proved by Leray \cite{Le} in $1934$, for 
$ \Om  = \mathbb{R}^3 $ and later, in $1951$, Hopf  \cite{Ho} solved the problem for a bounded open domain $ \Om  \subset \mathbb{R}^n , n \ge 2$, with homogeneous Dirichlet conditions on smooth boundaries, $\partial  \Om $. These results were subsequently extended to include functions ${\bf{f}}({\bf{x}},t) \in  \mathbb{L}^2 [(0,T)\,;\, \mathbb{V}[ \Om ]^{ - 1} ]$, where $\mathbb{V} [ \Om ]^{ - 1}  = \mathbb{V} [ \Om ]^*$ is the dual of $\mathbb{V}[ \Om]$ (see \cite{Li, T1, vW}).  In $1962$, Kato and Fujita \cite{KF} proved the existence of strong, global in time, smooth three-dimensional solutions, provided that the body forces are small (in an appropriate sense) and the initial data is small in  the Sobolev space $H^{1/2}[\Om]$ (see also \cite{CH} and Temam \cite[pages 205-208]{T1}).  (As noted by Temam \cite[see page 344-345]{T1}, the importance of their work is that it points out the dependence of the existence of global solutions on the size of the initial data, the body forces and possibly the spectral properties of these quantities.)  In another (related) direction,  Raugel and Sell showed that one can get stronger results for thin 3D domains (see \cite{RS}).  In this case, they show that the Navier-Stokes equations have strong solutions and that the long-time dynamics has a global attractor.   

For $n=3$, let $\mathbb{P}$ be the (Leray) orthogonal projection of $(\mathbb{L}^2 [ \Om ])^3$  onto $\mathbb{H}$
 and define the Stokes operator by: $ {\bf{Au}} = : -\mathbb{P} \Delta {\bf{u}}$, for ${\bf{u}} \in D({\bf{A}}) \subset \mathbb{H}^{2}[ \Om] \cap  {\mathbb{H}}_0^{1}[ \Om]$, the domain of ${\bf{A}}$.  The purpose of this paper is to prove that there exists a number $ {{{u}}_ +} $, depending only on ${\mathbf{A}}$,  $f$,  $\nu $ and $ \Om $, such that, for all functions in 
$
\mathbb{D} = D({\mathbf{A}}) \cap \mathbb{B},
$
 where $D({\mathbf{A}})$ is the domain of ${\mathbf{A}}$ and
${{\mathbb B}}$ is the closed ball of radius 
$ \tfrac{1}{2}{{u}_ +} $, in ${{\mathbb H}}$, the Navier-Stokes equations have unique, strong, solutions in 
$
{\bf{u}}\in L_{\text{loc}}^\infty[[0,\infty); {\mathbb V}( \Om )]
\cap \mathbb{C}^1[(0,\infty);{\mathbb H}].
$
\section*{Preliminaries}
Applying the Leray projection to equation (1), with 
${{B}}({\mathbf{u}},{\mathbf{u}}) = \mathbb{P}({\mathbf{u}} \cdot \nabla ){\mathbf{u}}$, we can recast equation (1) in the standard form:
\beqn
\begin{gathered}
  \partial _t {\mathbf{u}} =  - \nu {\mathbf{Au}} - {{B}}({\mathbf{u}},{\mathbf{u}}) + \mathbb{P}{\mathbf{f}}(t){\text{ in (}}0,T) \times \Omega , \hfill \\
  {\text{                              }}{\mathbf{u}}(t,{\mathbf{x}}) = {\mathbf{0}}{\text{ on (}}0,T) \times \partial \Omega,\;\;{\mathbf{u}}(0,{\mathbf{x}}) = {\mathbf{u}}_0 ({\mathbf{x}}){\text{ in }}\Omega, \hfill \\ 
\end{gathered} 
\eeqn
where we have used the fact that the orthogonal complement of ${\Ha}$ relative to $({L}^{2} [\Omega ])^3 $ is $\{ {\mathbf{v}}\,:\;{\mathbf{v}} = \nabla q,\;q \in (H^1 [\Omega ])^3 \} $ to eliminate the pressure term (see Galdi [GA] or [SY, T1,T2]). 
\begin{Def}  We say that the operator ${\mathbf{J}}( \cdot ,t)$ is (for each $t$) 
\begin{enumerate}
\item
0-Dissipative if $
\left\langle {{\mathbf{J}}({\mathbf{u}},t),{\mathbf{u}}} \right\rangle _{\mathbb{H}}  \le 0$.
\item
Dissipative if 
$\left\langle {{\mathbf{J}}({\mathbf{u}},t) - {\mathbf{J}}({\mathbf{v}},t),{\mathbf{u}} - {\mathbf{v}}} \right\rangle _{\mathbb{H}}  \le 0$.
\item
Strongly dissipative if there  exists an $ \alpha > 0$ such that
$$
\left\langle {{\mathbf{J}}({\mathbf{u}},t) - {\mathbf{J}}({\mathbf{v}},t),{\mathbf{u}} - {\mathbf{v}}} \right\rangle _{\mathbb{H}}  \le  - \alpha \left\| {{\mathbf{u}} - {\mathbf{v}}} \right\|_{\mathbb{H}}^2. 
$$
\end{enumerate}
\end{Def}
Note that, if ${\mathbf{J}}( \cdot ,t)$ is a linear operator, definitions (1) and (2) coincide.  Theorem 2 below is essentially due to Browder \cite{B},  see Zeidler {\cite[Corollary 32.27, page 868 and Corollary 32.35 page 887, in Vol. IIB]{Z}}, while Theorem 3 is from Miyadera \cite[p. 185, Theorem 6.20]{M}, and is a modification of the Crandall-Liggett Theorem \cite{CL} (see the appendix to the first section of \cite{CL}) . 
\begin{thm} Let $\mathbb{B}$ be a closed, bounded, convex subset of $
\mathbb{H}$.  If ${\mathbf{J}}( \cdot ,t):\mathbb{B}[ \Om ] \to \mathbb{H}$ is strongly dissipative for each fixed $t \ge 0$, then for each ${\mathbf{b}} \in \mathbb{B}$, there is a 
${\mathbf{u}} \in \mathbb{B}$ with ${\mathbf{J}}({\mathbf{u}},t) = {\mathbf{b}}$ (i.e., the range, $
Ran{\text{[}}{\mathbf{J}}( \cdot ,t)] \supset \mathbb{B}$).
\end{thm}	
\begin{thm} Let ${\text{\{ }}\mathcal{A}(t), t \in I = [0,\infty ){\text{\} }}$ be a family of operators defined on $\mathbb{H}$ with domains $
D(\mathcal{A}(t)) = D$, independent of $t$. (We assume that the closure of $\mathbb{D} = D(A)\cap {\mathbb{B}}$, in the $\mathbb H$ norm equals  ${\mathbb{B}}$):
\begin{enumerate}
\item
The operator $\mathcal{A}(t)$ is the generator of a contraction semigroup for each
$t \in I$.
\item
The function $\mathcal{A}(t){\mathbf{u}}$ is continuous in both variables on $
I \times \mathbb{D}$.
\end{enumerate}
Then, for every ${\mathbf{u}}_0  \in \mathbb{D}$, the problem 
$\partial _t {\mathbf{u}}(t,{\mathbf{x}}) = \mathcal{A}(t){\mathbf{u}}(t,{\mathbf{x}})$, 
${\mathbf{u}}(0,{\mathbf{x}}) = {\mathbf{u}}_0 ({\mathbf{x}})$, has a unique solution 
${\mathbf{u}}(t,{\mathbf{x}}) \in \mathbb{C}^1 (I;\mathbb{H})$.
\end{thm} 
\subsection*{Stokes Equation}
The difficulty in proving the existence of global-in-time strong solutions for equation (2) can be directly linked to the problem of getting good estimates for the nonlinear term ${{B}}({\mathbf{u}},{\mathbf{u}})$.  For example, the following theorem is one of the major estimates used to study this equation (see equation 61.22 on page 366, in  Sell and You \cite{SY} and Constantin and Foias \cite{CF}).  (We assume that ${\bf{u}},\, {\bf{v}} \in D(A)$.)
 
\begin{thm} Let $ \Om $ be a bounded open set of class $\mathbb{C}^k$ in $\mathbb{R}^3$.  Let ${\alpha_i,1 \le i \le 3}$, satisfy 
$ {0 \le \alpha_1\le k}$, ${0 \le \alpha_2 \le k-1}$, ${0 \le \alpha_3 \leq k}$, with ${ \alpha_1+\alpha_2+\alpha_3 \ge 3/2}$ and
\beqa
(\alpha _1 ,\alpha _2 ,\alpha _3 ) \notin \left\{ {(3/2,0,0),(0,3/2,0),(0,0,3/2)} \right\}.
\eeqa
Then there is a positive constant $c=c(\al_i, \Om)$ such that
\beqn
\left| {\left\langle {{{B}}({\mathbf{u}},{\mathbf{v}}),{\mathbf{w}}} \right\rangle _\mathbb{H} } \right| \le c\left\| {{\mathbf{A}}^{\alpha _1 /2} {\mathbf{u}}} \right\|_\mathbb{H} \left\| {{\mathbf{A}}^{(1 + \alpha _2 )/2} {\mathbf{v}}} \right\|_\mathbb{H} \left\| {{\mathbf{A}}^{\alpha _3 /2} {\mathbf{w}}} \right\|_\mathbb{H}. 
\eeqn
\end{thm}
We plan to show that, by renorming $\mathbb{H}$, we can prove a very strong inequality for equation (3).  First we need to investigate  the Stokes equation.

If we drop the nonlinear term, we get the well-known Stoke's equation ($\mathbb{P}{\mathbf{f}}(t)={\mathbf 0}$):
\beqa
\begin{gathered}
  \partial _t {\mathbf{u}} =  - \nu {\mathbf{Au}}  {\text{ in (}}0,T) \times \Omega , \hfill \\
  {\text{                              }}{\mathbf{u}}(t,{\mathbf{x}}) = {\mathbf{0}}{\text{ on (}}0,T) \times \partial \Omega , \hfill \\
  {\text{                              }}{\mathbf{u}}(0,{\mathbf{x}}) = {\mathbf{u}}_0 ({\mathbf{x}}){\text{ in }}\Omega. \hfill \\ 
\end{gathered} 
\eeqa
A proof of the next theorem holds and may be found in Sell and You \cite{SY} (page 114):
\begin{thm}
Let $\Om$ be a open bounded domain of class ${\mathbb{C}}^2$ in  ${\mathbb{R}}^3$, and let $\bf{A}$ be the Stokes operator on $\Om$.  Then the following holds:
\begin{enumerate}
\item The operator $\bf{A}$  is a positive selfadjoint generator of a contraction semigroup $S(t)$. 
\item The inverse of ${\bf{A}},\;{\bf{A}}^{-1}$, is a compact linear operator from $\mathbb{H}$ onto $D({\bf{A}})$.
\item The operator $\bf{A}$ is sectorial and there exist eigenvalues $0 < \lambda_1 \le \lambda_2 \le \cdots,\; \lambda_n \rightarrow \infty$, as $n \rightarrow \infty$ and, $\left\| {S(t)} \right\| \le e^{-{\lambda_1}t}$.  
\item The eigenfunctions, $\{ {\bf{e}}_1,\ {\bf{e}}_2,\ \cdots \}$, of $\bf{A}$ form an orthonormal basis for $\mathbb{H}$.
\end{enumerate}
\end{thm}
\subsection*{Equivalent Norms}
\begin{ex}
In order to see how we can use the analytic properties of a semigroup to bound the generator,  let $\mathcal{H} = {L}^2 [\mathbb{R}^3, d{\mu}]$, where $d{\mu}=(2\pi)^{-3/2}e^{-\frac{1}{2}\left|{\bf{x}}\right|^2}d{\bf{x}}$, and consider the problem:
\[
\frac{\partial }
{{\partial t}}u(t,{\mathbf{x}}) = \Delta u(t,{\mathbf{x}})-{\bf{x}} \cdot \nabla {\bf{u}}({\bf{x}},t),\quad u(0,{\mathbf{x}}) = u_0 ({\mathbf{x}}).
\]
This is the well-known Ornstein-Uhlenbeck equation with solution $(T(t)u_0) ({\mathbf{x}}) = u(t,{\mathbf{x}})$, where:
\[
(T(t){\bf{u}}_0 )({\bf{x}}) = \frac{1}
{{\sqrt {\left[ {2\pi (1 - e^{ - t} )} \right]^3 } }}\int_{\mathbb{R}^{3} } {exp\left\{ { { - }\frac{{\left( {e^{ - t/2} {\bf{x}} - {\bf{y}}} \right)^2 }}
{{2 (1 - e^{ - t} )}}} \right\}{\bf{u}}_0 ({\bf{y}})d{\bf{y}}}. 
\]
The operator $T(t)$ is a (analytic) transition semigroup,  $T(t){\mathbf{1}}={\mathbf{1}}$, with generator
 $D^2=\Delta-{\bf{x}} \cdot \nabla$. For general initial data ${u_0}$, there is a positive number $\om$  such that  $\left\| {T(t)u_0 } \right\|_2  \le e^{-\om t} \left\| {u_0 } \right\|_2 $.  If we set $S(t)= e^{\om t}T(t)$, then $S(t)$ is a uniformly bounded  analytic semigroup, $\left\| {S(t)u_0 } \right\|_2  \le  \left\| {u_0 } \right\|_2$.  It follows that there is a positive constant $M$ such that for any fixed $t$ 
\[
\left\| {S(t)u_0 } \right\|_2  \le  \left\| {u_0 } \right\|_2 \le M \left\| {S(t)u_0 } \right\|_2.
\] 
Thus, if we set $\left\| w \right\|_{2,{\gamma}}   = \left\|{S(\gamma )w} \right\|_2 \le  \left\| w \right\|_2$ for any $\gamma  \in (0,1)$, we obtain  an equivalent norm on $\mathcal{H}$.  Moreover, since $S(t)$ is analytic, if $w$ is in the domain of $D^2$, there is a constant $c > 0$ such that 
\[
\left\| {D^2 w} \right\|_{2,\gamma }  = e^{  \omega \gamma} \left\| {D^2 T(\gamma )w} \right\|_2  \leqslant e^{  \omega \gamma }  e^{-\omega \gamma } \frac{c}
{\gamma }\left\| w \right\|_2  \leqslant \frac{{Mc}}
{\gamma }\left\| w \right\|_{2,\gamma }.
\] 
We thus conclude that this equivalent norm on $\mathcal{H}$ makes $D^2$ bounded (on its domain) without using the graph norm.  Note that $\Delta= D^2+{\bf{x}} \cdot \nabla$.  From here, it is easy to see that the equivalent norm also makes $\Delta$ and $\nabla$  bounded in the above sense.  In this latter case, we obtain the reverse of the Poincar\'{e} inequality.  
\end{ex}
In our case, we let $T(t)=exp\{-t\bf{A}\}$ be the analytic semigroup  generated by the Stokes operator $\bf{A}$, with $\left\| {T(t)\bf u} \right\|_\Ha  \leqslant  e^{-\omega t} \left\| \bf u \right\|_\Ha$.  Let $S(t)= e^{\om t}T(t)$ and choose $M$ as in our example, so that $\left\| \bf u \right\|_{ {\Ha},1}  = \left\| {S(r)\bf u} \right\|_ {\mathbb{H}}$ is an equivalent norm, where $r$ is to be determined.  Since $\bf{A}$ is analytic,  there is a constant $c_z$ such that, for ${\bf u} \in D({\bf A^z})$,
\[
\left\| {{\mathbf{A}}^z {\mathbf{u}}} \right\|_{\mathbb{H},1}  = e^{ \omega r} \left\| {{\mathbf{A}}^z T(r){\mathbf{u}}} \right\|_\mathbb{H}  \leqslant e^{ \omega r}  e^{-\omega r} \frac{c}
{{(r)^z }}\left\| {\mathbf{u}} \right\|_\mathbb{H}  \leqslant \frac{{Mc}}
{{(r)^z }}\left\| {\mathbf{u}} \right\|_{\mathbb{H},1}.
\]
Since the norms are equivalent, we also have (for ${\bf u} \in D({\bf A^z})$):
\[
\left\| {{\mathbf{A}}^z {\mathbf{u}}} \right\|_{\mathbb{H}}  \le M \left\| {{\mathbf{A}}^z {\mathbf{u}}} \right\|_{\mathbb{H},1} = e^{  \omega r} \left\| {{\mathbf{A}}^z T(r){\mathbf{u}}} \right\|_\mathbb{H}  \le  \frac{Mc}
{{(r)^z }}\left\| {\mathbf{u}} \right\|_\mathbb{H}.
\]
From Theorem 4, we have the following result: 
\begin{thm}
Let ${\bf u} \in D({\bf{A}})$, set ${\bf S}=S(r)$ and renorm $\mathbb{H}$ so that $\left\| \bf u \right\|_{\mathbb{H},1}  = \left\| {{\mathbf{S}} \bf u} \right\|_\mathbb{H}$.  Then, with $k=2$:
\begin{enumerate}
\item
If we let  ${\alpha_1=0}$, ${ \alpha_2=1}$ and $\alpha_3=1/2$, there are positive constants $c=c(\al_i, \Om)$ and $c_1$  such that
\beqn
\left| {\left\langle {{{\bf{A}}^{-1}}{{B}}({\mathbf{u}},{\mathbf{v}}),{\mathbf{w}}} \right\rangle _{\mathbb{H},1} } \right| \le \f{M^3cc_1}{r^{1/4}} \left\| {\mathbf{u}} \right\|_{\mathbb{H},1} \left\| {\mathbf{w}} \right\|_{\mathbb{H},1} \left\| {\mathbf{v}} \right\|_{\mathbb{H},1}. 
\eeqn
\item
If we let  ${\alpha_1=0}$, ${ \alpha_2=1}$ and $\alpha_3=1/2$, there are positive constants $c=c(\al_i, \Om)\; c_1$ and $c_2$  such that
\beqn
\left| {\left\langle {{{B}}({\mathbf{u}},{\mathbf{v}}),{\mathbf{w}}} \right\rangle _{\mathbb{H},1} } \right| \le\f{M^4c{ c_1}c_2}{r^{5/4}} \left\| {{\mathbf{u}}} \right\|_{\mathbb{H},1} \left\| { {\mathbf{v}}} \right\|_{\mathbb{H},1} \left\| { {\mathbf{w}}} \right\|_{\mathbb{H},1}. 
\eeqn
\end{enumerate}
\end{thm}
\begin{proof}
To prove (4), first note that ${\bf{A}} $ and $\bf{S}$ commute on D({\bf{A}}) so, if we set  ${\bf{S}}{\bf{w}}={\bf{w}}_1$, we have:  
\[
b( {{\bf{A}}^{-1}}{\mathbf{u}},{\mathbf{v}} ,{\mathbf{w}})_{\mathbb{H},1}  = \left\langle {{{\bf{A}}^{-1}}{\mathbf{SB}}({\mathbf{u}},{\mathbf{v}}),{\mathbf{Sw}}} \right\rangle _\mathbb{H}  = b({\mathbf{u}},{\mathbf{v}},{{\bf{S}}{\bf{A}}^{-1}}{\mathbf{w}}_1 )_{\mathbb{H}}. 
\]
Using the selfadjoint property of ${\bf{A}}$, and integration by parts, we have
$$
b({\bf{u}},{\bf{v}},{{\bf{S}}{\bf{A}}^{-1}}{\mathbf{w}}_1)_{\mathbb{H}}   =  -  b({\bf{u}},{{\bf{S}}{\bf{A}}^{-1}}{\mathbf{w}}_1,{\bf{v}})_{\mathbb{H}}.
$$
It now follows from Theorem 4 that:
$$
\left| {\left\langle {\bf{A}}^{-1}{{B}}({\mathbf{u}},{\mathbf{v}}),{\mathbf{w}} \right\rangle _{\mathbb{H},1}} \right| 
\le c\left\| {{\mathbf{A}}^{\alpha _1 /2} {\mathbf{u}}} \right\|_{\mathbb{H}} \left\| {{\bf S}{\mathbf{A}}^{ (1 + \alpha _2 )/2} {\bf{A}^{-1}}{\mathbf{w}}_{1} }\right\|_{\mathbb{H}} 
\left\| {{\mathbf{A}}^{\alpha _3 /2} {\mathbf{v}}} \right\|_{\mathbb{H}}. 
$$
If we set $ \alpha _1 = 0$, $\alpha _2 = 1$ and $\alpha _3 = 1/2$
we have 
\[
\begin{gathered}
  \left| {\left\langle {{\bf{A}}^{-1}{{B}}({\mathbf{u}},{\mathbf{v}}),{\mathbf{w}}} \right\rangle _{\mathbb{H}, 1} } \right| \le c\left\| {\mathbf{u}} \right\|_{\mathbb{H}} \left\| {{\mathbf{A}} {\bf{A}^{-1}}{\mathbf{w}}_1 } \right\|_\mathbb{H} \left\| {\bf{A}}^{1/4}{\bf{v}} \right\|_{\mathbb{H}} \hfill \\
  {\text{                       }} \le \f{Mcc_1}{r^{1/4}}\left\| {\mathbf{u}} \right\|_{\mathbb{H}} \left\| {\mathbf{w}}_1 \right\|_\mathbb{H} \left\| {\bf{v}} \right\|_{\mathbb{H}}  \hfill \\
  {\text{                       }} \le  \f{M^3cc_1}{r^{1/4}} \left\| {\mathbf{u}} \right\|_{\mathbb{H},1} \left\| {\mathbf{w}} \right\|_{\mathbb{H},1} \left\| {\mathbf{v}} \right\|_{\mathbb{H},1}.  \hfill \\ 
\end{gathered} 
\]

To prove (5), as before, set ${\bf{S}}{\bf{w}}={\bf{w}}_1$, to obtain:  
\[
b( {\bf{u}},{\bf{v}} ,{\bf{w}})_{\mathbb{H},1}  = \left\langle {{\bf{S}}B({\bf{u}},{\bf{v}}),{\bf{S}}{\bf{w}}} \right\rangle _\mathbb{H} = b({\bf{u}},{\bf{v}},{\bf{S}}{\bf{w}}_1 )_{\mathbb{H}}. 
\]
As before, using the selfadjoint property of ${\bf{A}}$, and integration by parts, we have
$$
b({\bf{u}},{\bf{v}},{\bf{S}}{\bf{w}}_1)_{\mathbb{H}}   =  -  b({\bf{u}},{\bf{S}}{\bf{w_{1}}},{\bf{v}})_{\mathbb{H}}.
$$
It follows that:
\beqn
\left| {\left\langle {{B}}({\bf{u}},{\bf{v}}),{\bf{w}} \right\rangle _{\mathbb{H},1}} \right| 
\le c\left\| {{\bf{A}}^{\alpha _1 /2} {\bf{u}}} \right\|_{\mathbb{H}} \left\| {{\bf{S}}{\bf{A}}^{ (1 + \alpha _2 )/2} {\bf{w}}_{1} }\right\|_{\mathbb{H}} 
\left\| {{\bf{A}}^{\alpha _3 /2} {\bf{v}}} \right\|_{\mathbb{H}}. 
\eeqn
Setting $ \alpha _1 =  0$, $\alpha _2 = 1$ and $\alpha _3 = 1/2$  we have: 
\[
\begin{gathered}
  \left| {\left\langle {{{B}}({\mathbf{u}},{\mathbf{v}}),{\mathbf{w}}} \right\rangle _{\mathbb{H},1} } \right| \le c\left\| {\bf{A}}^{1/4}{\bf{v}} \right\|_\mathbb{H} \left\| {{\mathbf{A}} {\mathbf{w}}_1 } \right\|_\mathbb{H} \left\| {\bf{u}} \right\|_{\mathbb{H}} \hfill \\
  {\text{                       }} \le \f{Mc{ c_1}c_2}{r^{5/4}} \left\| {\bf{u}} \right\|_{\mathbb{H}} \left\| {\bf{w}} \right\|_\mathbb{H} \left\| {\bf{v}} \right\|_{\mathbb{H}}  \hfill \\
  {\text{                       }} \le \f{M^4c{ c_1}c_2}{r^{5/4}} \left\| {\bf{u}} \right\|_{\mathbb{H},1} \left\| {\bf{w}} \right\|_{\mathbb{H},1} \left\| {\bf{v}} \right\|_{\mathbb{H},1}.  \hfill \\ 
\end{gathered} 
\]
\end{proof}
\begin{cor}
\[
max\{\left\| {{{B}}({\mathbf{u}},{\mathbf{v}})} \right\|_{\mathbb{H},1},\ \left\| {{{B}}({\mathbf{v}},{\mathbf{u}})} \right\|_{\mathbb{H},1} \} \leqslant \f{M^4c{ c_1}c_2}{r^{5/4}}\left\| {\mathbf{u}} \right\|_{\mathbb{H},1} \left\| {\mathbf{v}} \right\|_{\mathbb{H},1}. 
\]
\end{cor}
\section*{M-Dissipative Conditions} 
In the remainder of the paper, we assume that 
$
{\mathbf{f}}(t) \in L^\infty[[0,\infty); {\mathbb H}( \Om )]
$
and is H\"{o}lder continuous in $t$, with $\left\| {{\mathbf{f}}(t) - {\mathbf{f}}(\tau )} \right\|_{\mathbb{H},1}  \le d\left| {t - \tau } \right|^\theta,{\text{ }}d > 0,{\text{ }}0 < \theta  < 1$.  We can now rewrite equation (2) in the form:
\beqn
\begin{gathered}
  \partial _t {\mathbf{u}} = \nu {\mathbf{A}} {\mathbf{J}}({\mathbf{u}},t){\text{ in (}}0,T) \times \Omega , \hfill \\
  {\mathbf{J}}({\mathbf{u}},t) =  -  {\mathbf{u}} - \nu ^{ - 1} {\mathbf{A}}^{ - 1 } {{B}}({\mathbf{u}},{\mathbf{u}}) + \nu ^{ - 1} {\mathbf{A}}^{ - 1 } \mathbb{P}{\mathbf{f}}(t). \hfill \\ 
\end{gathered} 
\eeqn
\section*{Approach} 	
We begin with a study of the operator ${\mathbf{J}}( \cdot ,t)$, for fixed $t$, and seek conditions depending on ${\mathbf{A}},{\text{ }}\nu ,{\text{ }} \Om {\text{ and }}{\mathbf{f}}(t)$ which guarantee that ${\mathbf{J}}( \cdot ,t)$ is m-dissipative for each $t$.  Clearly $
{\mathbf{J}}( \cdot ,t):D({\mathbf{A}})\xrightarrow{{onto}}D({\mathbf{A}})$ and, since $ \nu {\mathbf{A}} =   \nu {\mathbb{P}}[-\Delta]$ is a closed positive (m-accretive) operator, so that $ - {\mathbf{A}}$ generates a linear contraction semigroup, we expect that $\nu {\mathbf{AJ}}( \cdot ,t)$ will be m-dissipative for each $t$.  
\begin{thm} For $t \in I=[0, \infty)$ and, for each fixed ${\mathbf{u}} \in D({\bf{A}}),\;
{\mathbf{J}}({\mathbf{u}},t)$ is H\"{o}lder continuous, with $
\left\| {{\mathbf{J}}({\mathbf{u}},t) - {\mathbf{J}}({\mathbf{u}},\tau )} \right\|_{\mathbb{H},1}  \le d'\left| {t - \tau } \right|^\theta$, where $d' = d{{\nu}^{-1}}( \lambda _1 )^{ -1} $, $d$ is the H\"{o}lder constant for the function ${\mathbf{f}}(t)$ and $\lambda _1 $ is the first eigenvalue of ${\mathbf{A}}$.
\end{thm}
\begin{proof}
For fixed ${\mathbf{u}} \in D({\bf{A}})$, 
\[
\begin{gathered}
  \left\| {{\mathbf{J}}({\mathbf{u}},t) - {\mathbf{J}}({\mathbf{u}},\tau )} \right\|_{\mathbb{H},1}  = \nu ^{ - 1} \left\| {{\mathbf{A}}^{ -1}{\bf{S}} {\text{[}}\mathbb{P}{\mathbf{f}}(t) - \mathbb{P}{\mathbf{f}}(\tau )]} \right\|_{\mathbb{H},1}  \hfill \\
  {\text{                                  }} \leq d{\nu^{-1}}( \lambda _1 )^{ -1} \left| {t - \tau } \right|^\theta   = d'\left| {t - \tau } \right|^\theta . \hfill \\ 
\end{gathered} 
\]
We have used the fact that every function $
{\text{ }}{\mathbf{h}}(t) \in \mathbb{H}( \Om )$ has an expansion in terms of the eigenfunctions of ${\mathbf{A}}$, so that $
{\mathbf{A}}^{ -1} {\text{ }}{\mathbf{h}}(t) = \sum\nolimits_{k = 1}^\infty  {\lambda _k^{ -1} h_k (t){\mathbf{e}}_k ({\mathbf{x}})}$ and, from here, it is easy to see that $
\left\| {{\mathbf{A}}^{ -1} {\text{ }}{\mathbf{h}}(t)} \right\|_{\mathbb{H},1}  \leq \lambda _1^{ -1} \left\| {{\text{ }}{\mathbf{h}}(t)} \right\|_{\mathbb{H},1} $.  
\end{proof}
\section*{Main Results} 
\begin{thm} Let $f = \sup _{t \in {\mathbf{R}}^ +  } \left\| {\mathbb{P}{\mathbf{f}}(t)} \right\|_{\mathbb{H},1}  < \infty $, then there exists a positive constant ${{u}}_ +  $, depending only on $f$, ${\mathbf{A}}$, $\nu $ and $ \Om $, such that for all ${\mathbf{u}}$, with $
\left\| {\mathbf{u}} \right\|_{\mathbb{H},1}  \le \tfrac{1}{2}{{u}_ +}$, ${\mathbf{J}}( \cdot ,t)$ is strongly dissipative. 
\end{thm}
\begin{proof} The proof of our first assertion has two parts. First, we require that the nonlinear operator ${\mathbf{J}}( \cdot ,t)$
 be 0-dissipative, which gives us an upper bound ${{u}}_ +  $, in terms of the norm (i.e., $\left\| {\mathbf{u}} \right\|_{\mathbb{H},1}  \le {{u}}_ + $ ).  We then use this part to show that ${\mathbf{J}}( \cdot ,t)$ is strongly dissipative on the closed ball, $ \mathbb{B} = \left\{ {{\mathbf{u}} \in D({\bf{A}}) : \left\| {\mathbf{u}} \right\|_{\mathbb{H},1}  \le {\tfrac{1}{2}{{u}_ +}}  } \right\}$.  

Part 1) 
From equation (7), we get that
\beqa
\begin{gathered}
  \left\langle {{\mathbf{J}}({\mathbf{u}},t),{\mathbf{u}}} \right\rangle _{\mathbb{H},1}  =  - \left\langle { {\mathbf{u}},{\mathbf{u}}} \right\rangle _{\mathbb{H},1}  + \nu ^{ - 1} \left\langle { - {\mathbf{A}}^{-1} {{B}}({\mathbf{u}},{\mathbf{u}}) + {\mathbf{A}}^{-1} \mathbb{P}{\mathbf{f}}(t),{\mathbf{u}}} \right\rangle _{\mathbb{H},1}  \hfill \\
 =  - \left\| { {\mathbf{u}}} \right\|_{\mathbb{H},1}^2  - \nu ^{ - 1} \left\langle {{\mathbf{A}}^{-1} {{B}}({\mathbf{u}},{\mathbf{u}}),{\mathbf{u}}} \right\rangle _{\mathbb{H},1}  + \nu ^{ - 1} \left\langle {{\mathbf{A}}^{-1} \mathbb{P}{\mathbf{f}}(t),{\mathbf{u}}} \right\rangle _{\mathbb{H},1}  \hfill \\
 =  - \left\| { {\mathbf{u}}} \right\|_{\mathbb{H},1}^2  + \nu ^{ - 1} \left\langle {{\mathbf{B}}({\mathbf{u}},{\mathbf{A}}^{-1} {\mathbf{u}}_1),{\mathbf{u}}} \right\rangle _{\mathbb{H}}  + \nu ^{ - 1} \left\langle {{\bf{A}}^{-1}\mathbb{P}{\mathbf{f}}(t), {\mathbf{u}}} \right\rangle _{\mathbb{H},1}  \hfill \\ 
\end{gathered}. 
\eeqa
It follows that
\beqa
\begin{gathered}
  \left\langle {{\mathbf{J}}({\mathbf{u}},t),{\mathbf{u}}} \right\rangle _{\mathbb{H},1}  \le  - \left\| { {\mathbf{u}}} \right\|_{\mathbb{H},1}^2  + \nu ^{ - 1} \left| {\left\langle {{\mathbf{B}}({\mathbf{u}},{\mathbf{A}}^{-1} {\mathbf{u}}_1),{\mathbf{u}}} \right\rangle _{\mathbb{H}} } \right| + \nu ^{ - 1} f\left\| {{\mathbf{A}}^{-1} {\mathbf{u}}} \right\|_{\mathbb{H},1}  \hfill \\
 \le  - \left\| { {\mathbf{u}}} \right\|_{\mathbb{H},1}^2  + \nu^{ - 1} \f{M^3c{ c_1}}{r^{1/4}}  \left\| {\mathbf{u}} \right\|_{\mathbb{H},1}^3  + (\nu \lambda _1 )^{ - 1} f\left\| { {\mathbf{u}}} \right\|_{\mathbb{H},1} . \hfill \\ 
\end{gathered} 
\eeqa
In the last line, we used our estimate from Theorem 7.   Thus, ${\mathbf{J}}( \cdot ,t)$ will be 0-dissipative if 
\beqa
 - \ \left\| {\mathbf{u}} \right\|_{\mathbb{H},1}^{2}  + \nu^{ - 1}  \f{M^3c{ c_1}}{r^{1/4}} \left\| {\mathbf{u}} \right\|_{\mathbb{H},1}^3  + (\nu \lambda _1 )^{ - 1} f\left\| {\mathbf{u}} \right\|_{\mathbb{H},1}  \le 0,
\eeqa
so that 
\beqa
\left\| {\mathbf{u}} \right\|_{\mathbb{H},1} \left[ {\nu^{ - 1}  \f{M^3c{ c_1}}{r^{1/4}} \left\| {\mathbf{u}} \right\|_{\mathbb{H},1}^2  -  \left\| {\mathbf{u}} \right\|_{\mathbb{H},1}  + (\nu \lambda _1 )^{ - 1} f} \right] \le 0.
\eeqa
Since $\left\| {\mathbf{u}} \right\|_{{\Ha},1}  > 0$, we have that ${\mathbf{J}}( \cdot ,t)$ is 0-dissipative if ($\delta  = \frac{{\nu r^{1/4}  }}
{{ M^3cc_1 }}$)
\beqa
(\delta)^{-1}\left\| {\mathbf{u}} \right\|_ {{\Ha},1}^2  -  \left\| {\mathbf{u}} \right\|_ {{\Ha},1}  + (\nu \lambda _1 )^{ - 1} f \leqslant 0.
\eeqa
Solving, we get that 
\beqa
{{u}}_ \pm   = \tfrac{1}
{2}\delta \left\{ {1 \pm \sqrt {1 - {{4f} \mathord{\left/
 {\vphantom {{4f} {(\delta \nu \lambda _1})}} \right.
 \kern-\nulldelimiterspace} {(\delta \nu \lambda _1} )}}}  \right\} = \tfrac{1}
{2}\delta \left\{ {1 \pm \sqrt {1 - \gamma } } \right\},
\eeqa
where $
\gamma  = \frac{{4M^3f cc_1 }}
{{\nu ^2 \lambda _1 }r^{1/4}}.
$
Since we want real distinct solutions, we must require that 
\beqa
\gamma  < 1 \Rightarrow \nu  > \left[ {\frac{{4M^3f cc_1 }}
{{r^{1/4}\lambda _1  }}} \right]^{1/2}.
\eeqa
It follows that, if $\mathbb{P}{\mathbf{f}} \ne {\mathbf{0}}$, then 
$
{{u}}_ -   < {{u}}_ + $ , and our requirement that ${\mathbf{J}}$
 is 0-dissipative implies that, since our solution factors as 
$
(\left\| {\mathbf{u}} \right\|_{{\Ha},1}  - {{u}}_ +  )(\left\| {\mathbf{u}} \right\|_{{\Ha},1}  - {{u}}_ -  ) \le 0,
$
we must have that:
\beqa
\left\| {\mathbf{u}} \right\|_{{\Ha},1}  - {{u}}_ +   \le 0,{\text{  }}\left\| {\mathbf{u}} \right\|_{{\Ha},1}  - {{u}}_-   \ge 0.
\eeqa
This means that whenever 
${\text{ }}{{u}}_ -   \le \left\| {\mathbf{u}} \right\|_{{\Ha},1}  \le {{u}}_ +  $, 
$
\left\langle {{\mathbf{J}}({\mathbf{u}},t),{\mathbf{u}}} \right\rangle _{{\Ha},1}  \le 0$.  (It is clear that when $
\mathbb{P}{\mathbf{f}}(t) = {\mathbf{0}}, {{u}}_ -   = {\mathbf{0}}$, and ${{u}}_ +   = \delta  $.)

Part 2): Now, for any  ${\mathbf{u}},{\mathbf{v}} \in \mathbb{D}$ with ${\mathbf{u}}-{\mathbf{v}} \in \mathbb{D}$ and $\max ({\text{ }}\left\| {\mathbf{u}} \right\|_{\mathbb{H},1} ,\left\| {\mathbf{v}} \right\|_{\mathbb{H},1} ) \le (1/2){{{u}}_ +}  $, we have that   
\beqa
\begin{gathered}
  \left\langle {{\mathbf{J}}({\mathbf{u}},t) - {\mathbf{J}}({\mathbf{v}},t),{\mathbf{u}} - {\mathbf{v}}} \right\rangle _{\mathbb{H},1}  =  - \left\| { ({\mathbf{u}} - {\mathbf{v}})} \right\|_{\mathbb{H},1}^2  \hfill \\
  {\text{                                                    }} - {\nu ^{ - 1}} \left\langle {{\mathbf{A}}^{-1} [{{B}}({\mathbf{u}} ,{\mathbf{u}}- {\mathbf{v}}) + {{B}}({\mathbf{u-v}}, {\mathbf{v}})],({\mathbf{u}} - {\mathbf{v}})} \right\rangle _{\mathbb{H},1}  \hfill \\
  {\text{                    }} \le  -  \left\| {{\mathbf{u}} - {\mathbf{v}}} \right\|_{\mathbb{H},1}^2  +  \nu^{ - 1} [1/(r^{1/4})]M^3c{ c_1} \left\| {{\mathbf{u}} - {\mathbf{v}}} \right\|_{\mathbb{H},1}^2 \left( {\left\| {\mathbf{u}} \right\|_{\mathbb{H},1}  + \left\| {\mathbf{v}} \right\|_{\mathbb{H},1} } \right) \hfill \\
  {\text{                    }} \le  - \left\| {{\mathbf{u}} - {\mathbf{v}}} \right\|_{\mathbb{H},1}^2  + \nu^{ - 1} [1/(r^{1/4})]M^3c{ c_1}\left\| {{\mathbf{u}} - {\mathbf{v}}} \right\|_{\mathbb{H},1}^2 {{u}}_ +   \hfill \\
  {\text{                    }} =  -  \left\| {{\mathbf{u}} - {\mathbf{v}}} \right\|_{{\Ha},1}^2  + \nu^{ - 1} [1/(r^{1/4})]M^3c{ c_1} \left\| {{\mathbf{u}} - {\mathbf{v}}} \right\|_{{\Ha},1}^2 \left[ \tfrac{1}
{2}{\delta \left\{ {1+\sqrt {1 - \gamma } } \right\}} \right] \hfill \\
  {\text{                    }} =  -  \tfrac{1}{2} \left\| {{\mathbf{u}} - {\mathbf{v}}} \right\|_{{\Ha},1}^2 \left\{ {1 - \sqrt {1 - \gamma } } \right\} \hfill \\
  {\text{                    }} =  - \alpha \left\| {{\mathbf{u}} - {\mathbf{v}}} \right\|_{{\Ha},1}^2 ,{\text{   }} \alpha = \tfrac{1}
{2} \left\{ {1 - \sqrt {1 - \gamma } } \right\}. \hfill \\ 
\end{gathered} 
\eeqa
\end{proof} 
\begin{thm} The operator ${\mathcal{A}}(t) = \nu {\mathbf{A}} \mathbf{J}( \cdot ,t)$
 is closed, strongly dissipative and jointly continuous in ${\mathbf{u}}$ and $t$.  Furthermore, for each $t \in {\mathbf{R}}^ +  $ and $\omega  > 0$, 
$Ran[I - \omega  {\mathcal{A}}(t)] \supset \mathbb{B}$, so that $
 {\mathcal{A}}(t)$ is m-dissipative on $\mathbb{D}$. 
\end{thm}
\begin{proof} 
  Since ${\mathbf{J}}( \cdot ,t)$ is strongly dissipative on $
\mathbb{B}[ \Om ]$, it follows from Theorem 2 that $
Ran[{\mathbf{J}}( \cdot ,t)] \supset \mathbb{B}$.

To show that ${\mathcal{A}}(t) = \nu {\mathbf{A}} {\mathbf{J}}( \cdot ,t)$ is strongly dissipative, for $
{\mathbf{u}},{\mathbf{v}} \in \mathbb{B}$, we have
\beqa
\left\langle {{\mathcal{A}}(t){\mathbf{u}} - {\mathcal{A}}(t) {\mathbf{v}} , ({\mathbf{u}} - {\mathbf{v}})} \right\rangle _{\mathbb{H},1}  =  - \nu \left\| {{\mathbf{A}}^{1/2} ({\mathbf{u}} - {\mathbf{v}})} \right\|_{\mathbb{H},1}^2  - \left\langle {[{{B}}({\mathbf{u}} ,{\mathbf{u}}- {\mathbf{v}}) + {{B}}({\mathbf{u}} - {\mathbf{v}},{\mathbf{v}})],({\mathbf{u}} - {\mathbf{v}})} \right\rangle _{\mathbb{H},1}  
\eeqa
Now, from equation (5) and Corollary 8, 
\beqa
\left| {\left\langle {[{{B}}({\mathbf{u}} ,{\mathbf{u}}- {\mathbf{v}}) + {{B}}({\mathbf{u}} - {\mathbf{v}},{\mathbf{v}})],({\mathbf{u}} - {\mathbf{v}})} \right\rangle _{\mathbb{H},1} } \right| \le \left[ {\frac{{M^4 cc_1 c_2 }}
{{r^{5/4} }}} \right]
\left\| {{\mathbf{u}} - {\mathbf{v}}} \right\|_{\mathbb{H},1}^2   \left\{ {\left\| {{\mathbf{u}}} \right\|_{\mathbb{H},1}   + \left\| {{\mathbf{v}}} \right\|_{\mathbb{H},1}  } \right\}.
\eeqa
We now have that 
\beqa
\begin{gathered}
  \left\langle {{\mathcal{A}}(t){\mathbf{u}} - {\mathcal{A}}(t){\mathbf{v}},{\mathbf{u}} - {\mathbf{v}}} \right\rangle _{\mathbb{H}.1}  \le  - \nu \left\| {{\mathbf{A}}^{1/2} ({\mathbf{u}} - {\mathbf{v}})} \right\|_{\mathbb{H},1}^2  + \left[ {\frac{{M^4 cc_1 c_2 }}
{{r^{5/4} }}} \right]
\left\| {{\bf{u}} - {\bf{v}}} \right\|_{\mathbb{H},1}^2  \left\{ {\left\| {{\mathbf{u}}} \right\|_{\mathbb{H},1}  + \left\| {{\mathbf{v}}} \right\|_{\mathbb{H},1} } \right\} \hfill \\
  {\text{                                   }} \le  { - \nu \lambda_1 \left\| { {\mathbf{u}} - {\mathbf{v}}} \right\|_{\mathbb{H},1}^2  + \left[ {\frac{{M^4 cc_1 c_2 }}
{{r^{5/4} }}} \right]
\left\| {{\mathbf{u}} - {\mathbf{v}}} \right\|_{\mathbb{H},1}^2 \left[ {\left\| {{\mathbf{u}}} \right\|_{\mathbb{H},1}  + \left\| {{\mathbf{v}}} \right\|_{\mathbb{H}.1} } \right]}  \hfill \\
  {\text{                                   }} \le  { - \nu \lambda_1 \left\| { {\mathbf{u}} - {\mathbf{v}}} \right\|_{\mathbb{H},1}^2  +\left[ {\frac{M^4cc_1 c_2 }
{{r^{5/4} }}} \right]
\left\| {{\mathbf{u}} - {\mathbf{v}}} \right\|_{\mathbb{H},1}^2 {{u}}_+}   \hfill \\
  {\text{                                    }} \le  \left\| {{\mathbf{u}} - {\mathbf{v}}} \right\|_{\mathbb{H},1}^2 \left\{ { - \nu \lambda _1  + \tfrac{1}
{2}\nu  {\frac{{c_2 }}
{{r  }}} 
\left[ {1 + \sqrt {1 - \gamma } } \right]} \right\}. \hfill \\ 
\end{gathered} 
\eeqa 
Thus, if we set $r=\hat{r}=c_2/ \lambda_1$, we can set $
a =  \nu \lambda _1\left[ {1 - \sqrt {1 - \gamma } } \right]$,  so that
\[
\left\langle {{\mathcal{A}}(t){\mathbf{u}} - {\mathcal{A}}(t){\mathbf{v}},{\mathbf{u}} - {\mathbf{v}}} \right\rangle _{\mathbb{H},1}  \leqslant  - a \left\| {({\mathbf{u}} - {\mathbf{v}})} \right\|_{\mathbb{H},1}^2 .
\] 
It follows that $ {\mathcal{A}}(t)$ is strongly dissipative.  Since $ - {\mathbf{A}}$ is m-dissipative, for $\omega  > 0$, $Ran(I + \omega {\mathbf{A}}) = {\mathbb{H}}$.  As ${\mathbf{J}}$ is strongly dissipative, with $
Ran[{\mathbf{J}}] \supset \mathbb{B}$, and
${\mathbf{J}}( \cdot ,t):\mathbb{D}\xrightarrow{{onto}}\mathbb{D}$, 
$ {\mathcal{A}}(t)$ is maximal dissipative, and hence closed, so that 
$Ran[I - \omega  {\mathcal{A}}(t)] \supset \mathbb{B}[ \Om ]$.  It follows that $ {\mathcal{A}}(t)$ is m-dissipative on $\mathbb{B}$ for each $t \in {\mathbf{R}}^ +  $ (since $\mathbb{H}$ is a Hilbert space).
To see that $ {\mathcal{A}}(t){\mathbf{u}}$ is continuous in both variables, let ${\mathbf{u}}_n ,{\mathbf{u}} \in \mathbb{B}$
, $\left\| {({\mathbf{u}}_n  - {\mathbf{u}})} \right\|_{\mathbb{H}}  \to 0$
, with $t_n ,t \in I$ and $t_n  \to t$.  Then (see Corollary 8) 
\beqa
\begin{gathered}
  \left\| { {\mathcal{A}}(t_n ){\mathbf{u}}_n  -  {\mathcal{A}}(t){\mathbf{u}}} \right\|_{\mathbb{H},1}  \leqslant \left\| { {\mathcal{A}}(t_n ){\mathbf{u}} -  {\mathcal{A}}(t){\mathbf{u}}} \right\|_{\mathbb{H},1}  + \left\| { {\mathcal{A}}(t_n ){\mathbf{u}}_n  -  {\mathcal{A}}(t_n ){\mathbf{u}}} \right\|_{\mathbb{H},1}  \hfill \\
   = \left\| {{\text{[}}\mathbb{P}{\mathbf{f}}(t_n ) - \mathbb{P}{\mathbf{f}}(t)]} \right\|_{\mathbb{H},1}  + \left\| {\nu {\mathbf{A}}({\mathbf{u}}_n  - {\mathbf{u}}) + [{{B}}({\mathbf{u}}_n  ,{\mathbf{u}}_n - {\mathbf{u}}) + {{B}}({\mathbf{u}},{\mathbf{u}}_n  - {\mathbf{u}})]} \right\|_{\mathbb{H},1}  \hfill \\
   \leqslant d\left| {t_n  - t} \right|^\theta   + \nu \left\| {{\mathbf{A}}({\mathbf{u}_n} - {\mathbf{u}})} \right\|_{\mathbb{H},1}  + \left\| {{{B}}({\mathbf{u}}_n ,{\mathbf{u}}_n  - {\mathbf{u}}) + {{B}}( {\mathbf{u}},{\mathbf{u}}_n  - {\mathbf{u}})} \right\|_{\mathbb{H},1}  \hfill \\
   \leqslant d\left| {t_n  - t} \right|^\theta   + \nu \frac{{ M c_3 }}
{{{\hat{r}}  }}\left\| {({\mathbf{u}}_n  - {\mathbf{u}})} \right\|_{\mathbb{H},1}  + \frac{{ M^4cc_1 c_2 }}
{{{\hat{r}}^{5/4} }} \left\| {({\mathbf{u}}_n  - {\mathbf{u}})} \right\|_{\mathbb{H},1} \left\{ {\left\| {{\mathbf{u}}_n } \right\|_{\mathbb{H},1}  + \left\| {{\mathbf{u}}} \right\|_{\mathbb{H},1} } \right\} \hfill \\
   \leqslant d\left| {t_n  - t} \right|^\theta   + \nu \frac{{ M c_3 }}
{{{\hat{r}} }}\left\| {({\mathbf{u}}_n  - {\mathbf{u}})} \right\|_{\mathbb{H},1}  + 2 \frac{{ M^4 cc_1 c_2 }}
{{{\hat{r}}^{5/4} }} \left\| {({\mathbf{u}}_n  - {\mathbf{u}})} \right\|_{\mathbb{H}} {{u}}_ +  . \hfill \\ 
\end{gathered} 
\eeqa
It follows that $ {\mathcal{A}}(t){\mathbf{u}}$ is continuous in both variables.
\end{proof}
When $\mathbf{f}=\mathbf{0}$, $\mathbb{B}$ is a ball about $\mathbf{0}$. Thus, we can  equip $\mathbb{B}$ with the closure of $\mathbb{D}$ in the $\mathbb H$ norm.  In this case, it follows that $\mathbb{B}$ is a closed, bounded, convex set, so that: 
\begin{thm} For each $T \in {\mathbf{R}}^ +$, $t \in (0,T)$ and ${\mathbf{u}}_0  \in \mathbb{D} \subset \mathbb{B}$, the global in time Navier-Stokes initial-value problem in $\Omega  \subset \mathbb{R}^3 :$
\beqn
\begin{gathered}
  \partial _t {\mathbf{u}} + ({\mathbf{u}} \cdot \nabla ){\mathbf{u}} - \nu \Delta {\mathbf{u}} + \nabla p = {\mathbf{0}}{\text{ in (}}0,T) \times \Omega , \hfill \\
  {\text{                              }}\nabla  \cdot {\mathbf{u}} = 0{\text{ in (}}0,T) \times \Omega , \hfill \\
  {\text{                              }}{\mathbf{u}}(t,{\mathbf{x}}) = {\mathbf{0}}{\text{ on (}}0,T) \times \partial \Omega , \hfill \\
  {\text{                              }}{\mathbf{u}}(0,{\mathbf{x}}) = {\mathbf{u}}_0 ({\mathbf{x}}){\text{ in }}\Omega . \hfill \\ 
\end{gathered} 
\eeqn
 has a unique strong solution ${\mathbf{u}}(t,{\mathbf{x}})$, which is in
 ${L_{\text{loc}}^2}[[0,\infty); {\mathbb {H}}]$ and in
$L_{\text{loc}}^\infty[[0,\infty); {\mathbb V}]
\cap \mathbb{C}^1[(0,\infty);{\mathbb H}]$.
\end{thm}
\begin{proof}
Theorem 3 allows us to conclude that when ${\mathbf{u}}_0  \in \mathbb{D}$, the initial value problem is solved and the solution ${\mathbf{u}}(t,{\mathbf{x}})$ is in $\mathbb{C}^1[(0,\infty);{\mathbb D}]$.  Since $\mathbb{D} \subset \mathbb{H}^{2}$, it follows that ${\mathbf{u}}(t,{\mathbf{x}})$ is also in $\mathbb{ V}$, for each $t>0$.  It is now clear that for any $T>0$,
\[
\int_0^T {\left\| {{\mathbf{u}}(t,{\mathbf{x}})} \right\|_{\mathbb{H}}^2 dt}  < \infty ,{\text{ and }}\sup _{0 < t < T} \left\| {{\mathbf{u}}(t,{\mathbf{x}})} \right\|_{\mathbb{V}}^2  < \infty .
\]
This gives our conclusion.
\end{proof}
When $\mathbf{f} \ne \mathbf{0}$, ${u}_{-} \ne {0}$.  Let $\Bbbk  = \left\{ {{\mathbf{u}}\;:\;\left\| {\mathbf{u}} \right\|_{\mathbb{H},1}  \leqslant {{u}}_ -  } \right\}$ and set $\mathbb{B}_ -   = \mathbb{B}   \cap \Bbbk ^c$, where $\Bbbk^c$ is the complement of $\Bbbk $.  Thus, we can take the closure of  $\mathbb{B}{_{-}} \cap \mathbb{D}$ in the $\mathbb H$ norm and use the largest closed convex set containing the initial data, inside this set.
\section*{Discussion}
 One interesting aspect of Theorem 12 is that it is impossible to restrict ourselves to a ball if the body forces are nonzero (e.g., the initial fluid velocity can never be zero).  This result is expected on physical grounds.   
 
It is known that if ${\mathbf{u}_0} \in \mathbb{V}$, and $\mathbf{f}(t)$ is $L^{\infty}[(0,\infty), \mathbb{H}]$ then there is a time $T> 0$ , such that a weak solution with this data is uniquely determined on any subinterval of $[0,T)$ (see Sell and You page 396, \cite{SY}).   Thus, we also have that: 
\begin{cor} For each $t \in {\mathbf{R}}^ + $ and $
{\mathbf{u}}_0  \in \mathbb{D} $ the Navier-Stokes initial-value problem in $ \Om  \subset \mathbb{R}^3 :$
\beqn
\begin{gathered}
  \partial _t {\mathbf{u}} + ({\mathbf{u}} \cdot \nabla ){\mathbf{u}} - \nu \Delta {\mathbf{u}} + \nabla p = {\mathbf{f}}(t){\text{ in (}}0,T) \times \Omega , \hfill \\
  {\text{                              }}\nabla  \cdot {\mathbf{u}} = 0{\text{ in (}}0,T) \times \Omega , \hfill \\
  {\text{                              }}{\mathbf{u}}(t,{\mathbf{x}}) = {\mathbf{0}}{\text{ on (}}0,T) \times \partial \Omega , \hfill \\
  {\text{                              }}{\mathbf{u}}(0,{\mathbf{x}}) = {\mathbf{u}}_0 ({\mathbf{x}}){\text{ in }}\Omega . \hfill \\ 
\end{gathered} 
\eeqn
 has a unique weak solution
${\mathbf{u}}(t,{\mathbf{x}})$, which is in
 ${L_{\text{loc}}^2}[[0,\infty); {\mathbb {H}}]$ and in
$L_{\text{loc}}^\infty[[0,\infty); {\mathbb V}]
\cap \mathbb{C}^1[(0,\infty);{\mathbb H}]$.
\end{cor}
Since we require that our initial data be in $\mathbb{H}^{2}$, the conditions for the Leray-Hopf weak solutions are not satisfied.  However, it was an open question as to whether these solutions developed singularities, even if ${\mathbf{u}}_0 \in \mathbb{C}_0^\iy $ (see Giga \cite{G}, and references therein).  The above Corollary shows that it suffices that 
${\mathbf{u}}_0 ({\mathbf{x}})\in {\mathbb {H}^2}$ to insure that the solutions develop no singularities.
\acknowledgements
We would like to thank Professor George Sell for introducing us to this problem and his help along the way.  We have benefited from his friendship, encouragement and the generous sharing of his knowledge over the last twenty years.  We would also like to sincerely thank Professor Edriss Titi for comments which helped us improve our presentation and eliminate a few areas of possible confusion.

\end{document}